\documentclass{article}
\usepackage{amssymb}
\usepackage{amsfonts}
\usepackage{amsmath}
\usepackage[doublespacing]{setspace}

\input{tcilatex}

\begin{document}

\title{Spherical-separablility of non-Hermitian Hamiltonians and pseudo-$%
\mathcal{PT}$-symmetry}
\author{Omar Mustafa$^{1}$ and S.Habib Mazharimousavi$^{2}$ \\
%EndAName
Department of Physics, Eastern Mediterranean University, \\
G Magusa, North Cyprus, Mersin 10,Turkey\\
$^{1}$E-mail: omar.mustafa@emu.edu.tr\\
$^{2}$E-mail: habib.mazhari@emu.edu.tr}
\maketitle

\begin{abstract}
Non-Hermitian but $\mathcal{P}_{\varphi }\mathcal{T}_{\varphi }$-symmetrized
spherically-separable Dirac and Schr\"{o}dinger Hamiltonians are considered.
It is observed that the descendant Hamiltonians $H_{r}$, $H_{\theta },$ and $%
H_{\varphi }$ play essential roles and offer some "user-feriendly" options
as to which one (or ones) of them is (or are) non-Hermitian. Considering a $%
\mathcal{P}_{\varphi }\mathcal{T}_{\varphi }$-symmetrized $H_{\varphi }$, we
have shown that the conventional Dirac (relativistic) and Schr\"{o}dinger
(non-relativistic) energy eigenvalues are recoverable. We have also
witnessed an unavoidable change in the azimuthal part of the general
wavefunction. Moreover, setting a possible interaction $V\left( \theta
\right) \neq 0$ in the descendant Hamiltonian $H_{\theta }$ would manifest a
change in the angular $\theta $-dependent part of the general solution too.
Whilst some $\mathcal{P}_{\varphi }\mathcal{T}_{\varphi }$-symmetrized $%
H_{\varphi }$ Hamiltonians are considered, a recipe to keep the regular
magnetic quantum number $m$, as defined in the regular traditional Hermitian
settings, is suggested. Hamiltonians possess properties similar to the $%
\mathcal{PT}$-symmetric ones (here the non-Hermitian $\mathcal{P}_{\varphi }%
\mathcal{T}_{\varphi }$-symmetric Hamiltonians) are nicknamed as \emph{%
pseudo-}$\mathcal{PT}$\emph{-symmetric.}


\end{abstract}

\section{Introduction}

In the search for the reality conditions on the energy spectra/eigenvalues
of non-Hermitian Hamiltonians [1-35], it is nowadays advocated (with no
doubts) that the orthodoxal mathematical Hermiticity requirement to ensure
the reality of the spectrum of a Hamiltonian is not only fragile but also
physically deemed remote, obscure and strongly unnecessary. A tentative
weakening of the Hermiticity condition through Bender's and Boettcher's [1] $%
\mathcal{PT}$-symmetric quantum mechanics (PTQM) (with $\mathcal{P}$
denoting parity and $\mathcal{T}$ \ time-reversal, a Hamiltonian $H$ is $%
\mathcal{PT}$-symmetric if it satisfies $\mathcal{PT}H\mathcal{PT}=H$) has
offered an alternative axiom that allows for the possibility of
non-Hermitian Hamiltonians.

Such a PTQM theory, nevertheless, has inspired intensive research on the
non-Hermitian Hamiltonians and led to the so-called pseudo-Hermitian
Hamiltonians (i.e., a pseudo-Hermitian Hamiltonian $H$ satisfies $\eta
\,H\,\eta ^{-1}=H^{\dagger }$ or $\eta \,H=H^{\dagger }\,\eta $, where $\eta 
$ is a Hermitian invertible linear operator and $(^{\dagger })$ denotes the
adjoint) by Mostafazadeh [16-21] which form a broader class of non-Hermitian
Hamiltonians with real spectra and encloses within those $\mathcal{PT}$%
-symmetric ones. Moreover, not restricting $\eta $ to be Hermitian (cf.,
e.g., Bagchi and Quesne [33]), and linear and/or invertible (cf., e.g.,
Solombrino [28], Fityo [29], and Mustafa and Mazharimousavi [30-32]) would
weaken pseudo-Hermiticity and lead to real spectra.

Based on the inspiring example, nevertheless, by Bender, Brody and Jones
[35] that $H=p^{2}+x^{2}+2x$ is a non-$\mathcal{PT}$-symmetric whereas a
simple amendment $H=p^{2}+\left( x+1\right) ^{2}-1$ (that leaves the
Hamiltonian invariant and allows parity to perform reflection about $x=-1$
rather than $x=0$) would consequently classify $H$ as $\mathcal{PT}$%
-symmetric (i.e., reflection need not necessarily be through the origin) and
promoting Znojil's understanding [34] of Bender's and Boettcher's PTQM
(i.e., $\mathcal{P}$ and $\mathcal{T}$ need not necessarily mean just the
parity and time reversal, respectively), we may introduce [36] a
time-reversal-like,%
\begin{equation}
\mathcal{T}_{\varphi }:\widehat{L}_{z}=-i\partial /\partial \varphi
\longrightarrow -\widehat{L}_{z}=i\partial /\partial \varphi ,\varphi
\longrightarrow \varphi ,i\longrightarrow -i
\end{equation}%
and a parity-like%
\begin{equation}
\mathcal{P}_{\varphi }:\widehat{L}_{z}\longrightarrow -\widehat{L}%
_{z},\varphi \longrightarrow \left( 2\pi -\varphi \right) ,
\end{equation}%
operators that might very well be accommodated by Bender's and Boettcher's
PTQM.

In this case, $\mathcal{P}_{\varphi }$ acting on a function $f\left(
r,\theta ,\varphi \right) \in L_{2}$ would read $\mathcal{P}_{\varphi
}f\left( r,\theta ,\varphi \right) =f\left( r,\theta ,2\pi -\varphi \right) $%
. Moreover, $f\left( r,\theta ,\varphi \right) $ is said to be $\mathcal{P}%
_{\varphi }\mathcal{T}_{\varphi }$-symmetric if it satisfies $\mathcal{P}%
_{\varphi }\mathcal{T}_{\varphi }f\left( r,\theta ,\varphi \right) =f\left(
r,\theta ,\varphi \right) $. Hence, our new operators leave the coordinates $%
r$ and $\theta $ unaffected and are designed to operate only on the
azimuthal descendent eigenvalue equation (e.g., equation (21) below) of the
spherically-separable non-Hermitian Hamiltonians (4) and (5). However, it
should be noted that our parity-like operator $\mathcal{P}_{\varphi }$ in
(2) is Hermitian, unitary, and performs reflection through a 2D-mirror
represented by the $xz$-plane. Yet, the proof of the reality of the
eigenvalues of a $\mathcal{P}_{\varphi }\mathcal{T}_{\varphi }$-symmetric
Hamiltonian is straightforward. Let the eigenvalue equation of our $\mathcal{%
P}_{\varphi }\mathcal{T}_{\varphi }$-symmetric Hamiltonian be $H\psi \left(
r,\theta ,\varphi \right) =E\psi \left( r,\theta ,\varphi \right) $, then $%
\mathcal{P}_{\varphi }\mathcal{T}_{\varphi }H\psi =\mathcal{P}_{\varphi }%
\mathcal{T}_{\varphi }E\psi =E\psi $. Using $\left[ \mathcal{P}_{\varphi }%
\mathcal{T}_{\varphi },H\right] =0$ we obtain $E\psi =E^{\ast }\psi $ and $E$
is therefore pure real (in analogy with Bender, Brody and Jones in [35] and
fits into PTQM-recipe).

In the forthcoming proposal, using spherical coordinates, we depart from the
traditional radial potential setting (i.e., $V\left( \mathbf{r}\right)
=V\left( r\right) $) into a more general potential of the the form%
\begin{equation}
V\left( \mathbf{r}\right) =V\left( r,\theta ,\varphi \right) =V\left(
r\right) +\left[ \frac{V\left( \theta \right) +V\left( \varphi \right) }{%
r^{2}\sin ^{2}\theta }\right] .
\end{equation}%
We shall use such potential setting in the context of Schr\"{o}dinger
Hamiltonian%
\begin{equation}
H=-\nabla ^{2}+V\left( \mathbf{r}\right) ,
\end{equation}%
and within an equally-mixed vector, $V\left( \mathbf{r}\right) $, and
scalar, $S\left( \mathbf{r}\right) $, potentials' setting in the Dirac
Hamiltonian%
\begin{equation}
H=\mathbf{\alpha \cdot p}+\beta \left[ M+S\left( \mathbf{r}\right) \right]
+V\left( \mathbf{r}\right) ,
\end{equation}%
with the possibility of non-Hermitian interactions' settings in the process.
However, it should be noted that such interactions in (3) with $V\left(
r\right) =-\alpha /r$, $V\left( \theta \right) =-b^{2}$, and $V\left(
\varphi \right) =0$ represent just variants of the well known Hartmann
potential [38-47] used in the studies of ring-shaped organic molecules.

For the sake of making our current proposal self-contained, we revisit, in
section 2, Dirac equation in spherical coordinates and give preliminary
foundation on its separability. We connect, in the same section, Dirac
descendant Hamiltonians with those of Schr\"{o}dinger and provide a clear
map for that. In section 3, we explore some consequences of a class of
complexified but $\mathcal{P}_{\varphi }\mathcal{T}_{\varphi }$-symmetrized
azimuthal Hamiltonians. For a complexified azimuthal interaction $V\left(
\varphi \right) \in 
%TCIMACRO{\U{2102} }%
%BeginExpansion
\mathbb{C}
%EndExpansion
$ (with $V\left( r\right) ,V\left( \theta \right) \in 
%TCIMACRO{\U{211d} }%
%BeginExpansion
\mathbb{R}
%EndExpansion
$) we use three illustrative examples for $V\left( \theta \right) =0$, $%
V\left( \theta \right) =1/2$, and $V\left( \theta \right) =1/\left( 2\cos
^{2}\theta \right) $. In section 4, a recipe of generating functions is
provided to keep the magnetic quantum number as is, whenever deemed
necessary of course. In the process of preserving the magnetic quantum
number $m$, a set of isospectral $\varphi $-dependent potentials, $V\left(
\varphi \right) $, for each set of $V\left( r\right) $ and $V\left( \theta
\right) $ is obtained. This would, moreover, allow reproduction of the
conventional-Hermitian relativistic and non-relativistic quantum mechanical
eigenvalues within our $\mathcal{P}_{\varphi }\mathcal{T}_{\varphi }$%
-symmetric non-Hermitian settings. We give our concluding remarks in section
5.

\section{Separability and preliminaries of Dirac and Schr\"{o}dinger
equations revisited}

Dirac equation with scalar and vector potentials, $S\left( \mathbf{r}\right) 
$ and $V\left( \mathbf{r}\right) $, respectively, reads (in $\hslash =c=1$
units)%
\begin{equation}
\left\{ \mathbf{\alpha \cdot p}+\beta \left[ M+S\left( \mathbf{r}\right) %
\right] +V\left( \mathbf{r}\right) \right\} \psi \left( \mathbf{r}\right)
=E\psi \left( \mathbf{r}\right) ,
\end{equation}%
where%
\begin{equation}
\mathbf{p=}-i\mathbf{\nabla }\text{ , \ }\mathbf{\alpha }=\left( 
\begin{array}{cc}
0 & \mathbf{\sigma } \\ 
\mathbf{\sigma } & 0%
\end{array}%
\right) \text{ , \ }\beta =\left( 
\begin{array}{cc}
I & 0 \\ 
0 & -I%
\end{array}%
\right) ,
\end{equation}%
and $\mathbf{\sigma }$ is the vector Pauli spin matrix. A Pauli-Dirac
representation would, with%
\begin{equation}
\psi \left( \mathbf{r}\right) =\left( 
\begin{array}{c}
\chi _{1}\left( \mathbf{r}\right) \\ 
\chi _{2}\left( \mathbf{r}\right)%
\end{array}%
\right) \text{,}
\end{equation}%
yield the decoupled equations%
\begin{equation}
\left( \mathbf{\sigma \cdot p}\right) \chi _{2}\left( \mathbf{r}\right) =%
\left[ E-V\left( \mathbf{r}\right) -M-S\left( \mathbf{r}\right) \right] \chi
_{1}\left( \mathbf{r}\right) ,
\end{equation}%
\begin{equation}
\left( \mathbf{\sigma \cdot p}\right) \chi _{1}\left( \mathbf{r}\right) =%
\left[ E-V\left( \mathbf{r}\right) +M+S\left( \mathbf{r}\right) \right] \chi
_{2}\left( \mathbf{r}\right) .
\end{equation}%
An equally-mixed scalar and vector potentials (i.e., $S\left( \mathbf{r}%
\right) =V\left( \mathbf{r}\right) $) leads to%
\begin{equation}
\chi _{2}\left( \mathbf{r}\right) =\frac{\mathbf{\sigma \cdot p}}{E+M},
\end{equation}%
and%
\begin{equation}
\left[ -\mathbf{\nabla }^{2}+2\left( E+M\right) V\left( \mathbf{r}\right) %
\right] \chi _{1}\left( \mathbf{r}\right) =\left[ E^{2}-M^{2}\right] \chi
_{1}\left( \mathbf{r}\right) .
\end{equation}%
Departing from the traditional "just-radially-symmetric" vector potential
(i.e., $V\left( \mathbf{r}\right) =V\left( r\right) $) into a more general,
though rather informative, vector potential (in the 3D spherical coordinates 
$r$, $\theta $, and $\varphi $) of the form%
\begin{equation}
V\left( \mathbf{r}\right) =V\left( r,\theta ,\varphi \right) =V\left(
r\right) +\left[ \frac{V\left( \theta \right) +V\left( \varphi \right) }{%
r^{2}\sin ^{2}\theta }\right] ,
\end{equation}%
would, with%
\begin{equation}
\chi _{1}\left( \mathbf{r}\right) =\chi _{1}\left( r,\theta ,\varphi \right)
=R\left( r\right) \Theta \left( \theta \right) \Phi \left( \varphi \right) ,
\end{equation}%
imply%
\begin{gather}
\frac{1}{R\left( r\right) }\left\{ \frac{\partial }{\partial r}\left( r^{2}%
\frac{\partial }{\partial r}\right) -2\left( E+M\right) V\left( r\right)
r^{2}+\left( E^{2}-M^{2}\right) r^{2}\right\} R\left( r\right)  \notag \\
+\frac{1}{\Theta \left( \theta \right) \sin \theta }\left[ \frac{\partial }{%
\partial \theta }\left( \sin \theta \frac{\partial }{\partial \theta }%
\right) -\frac{2\left( E+M\right) V\left( \theta \right) }{\sin \theta }%
\right] \Theta \left( \theta \right)  \notag \\
+\frac{1}{\Phi \left( \varphi \right) \sin ^{2}\theta }\left( \frac{\partial
^{2}}{\partial \varphi ^{2}}-2\left( E+M\right) V\left( \varphi \right)
\right) \Phi \left( \varphi \right) =0
\end{gather}%
The separability of which is obvious and mandates%
\begin{equation}
\left\{ \frac{1}{r^{2}}\frac{d}{dr}\left( r^{2}\frac{d}{dr}\right) -\frac{%
\Lambda }{r^{2}}-2\left( E+M\right) V\left( r\right) +\left(
E^{2}-M^{2}\right) \right\} R\left( r\right) =0,
\end{equation}%
\begin{equation}
\left[ \frac{1}{\sin \theta }\frac{d}{d\theta }\left( \sin \theta \frac{d}{%
d\theta }\right) -\left( \frac{2\left( E+M\right) V\left( \theta \right)
+m^{2}}{\sin ^{2}\theta }\right) +\Lambda \right] \Theta \left( \theta
\right) =0,
\end{equation}%
\begin{equation}
\left( \frac{d^{2}}{d\varphi ^{2}}-2\left( E+M\right) V\left( \varphi
\right) +m^{2}\right) \Phi \left( \varphi \right) =0,
\end{equation}%
where $m^{2}$ and $\Lambda $ are separation constants to be determined
below. Yet, in a straightforward manner, one can show that both Dirac and
Schr\"{o}dinger equations (with $V\left( \mathbf{r}\right) $ in (13)) would
read 
\begin{equation}
\left\{ \frac{1}{r^{2}}\frac{d}{dr}\left( r^{2}\frac{d}{dr}\right) -\frac{%
\Lambda }{r^{2}}-V_{eff}\left( r\right) +\lambda \right\} R\left( r\right)
=0,
\end{equation}%
\begin{equation}
\left[ \frac{1}{\sin \theta }\frac{d}{d\theta }\left( \sin \theta \frac{d}{%
d\theta }\right) -\left( \frac{V_{eff}\left( \theta \right) +m^{2}}{\sin
^{2}\theta }\right) +\Lambda \right] \Theta \left( \theta \right) =0,
\end{equation}%
\begin{equation}
\left( \frac{d^{2}}{d\varphi ^{2}}-V_{eff}\left( \varphi \right)
+m^{2}\right) \Phi \left( \varphi \right) =0,
\end{equation}%
where%
\begin{equation}
V_{eff}\left( r\right) =\left\{ 
\begin{tabular}{ll}
$V\left( r\right) $ & $\text{for Schr\"{o}dinger}$ \\ 
$2\left( E+M\right) V\left( r\right) $ & $\text{for Dirac}$%
\end{tabular}%
\right. ,
\end{equation}%
\begin{equation}
V_{eff}\left( \theta \right) =\left\{ 
\begin{tabular}{ll}
$V\left( \theta \right) $ & $\text{for Schr\"{o}dinger}$ \\ 
$2\left( E+M\right) V\left( \theta \right) $ & $\text{for Dirac}$%
\end{tabular}%
\right. ,
\end{equation}%
\begin{equation}
V_{eff}\left( \varphi \right) =\left\{ 
\begin{tabular}{ll}
$V\left( \varphi \right) $ & $\text{for Schr\"{o}dinger}$ \\ 
$2\left( E+M\right) V\left( \varphi \right) $ & $\text{for Dirac}$%
\end{tabular}%
\right. ,
\end{equation}%
\begin{equation}
\lambda =\left\{ 
\begin{tabular}{ll}
$E$ & $\text{for Schr\"{o}dinger}$ \\ 
$E^{2}-M^{2}$ & $\text{for Dirac}$%
\end{tabular}%
\right. .
\end{equation}%
The map between Schr\"{o}dinger and Dirac equations is clear, therefore.
Moreover, one can safely name three "new" descendant Hamiltonians and recast
the corresponding eigenvalue equations (with $\lambda =E$ for Schr\"{o}%
dinger and $\lambda =E^{2}-M^{2}$ for Dirac) as%
\begin{equation}
\begin{tabular}{ccc}
$H_{r}R\left( r\right) =\lambda R\left( r\right) ,$ & $H_{\theta }\Theta
\left( \theta \right) =\Lambda \Theta \left( \theta \right) ,$ & $H_{\varphi
}\Phi \left( \varphi \right) =m^{2}\Phi \left( \varphi \right) .$%
\end{tabular}%
\end{equation}

Of course it is a straightforward to work out the explicit forms of $H_{r}$, 
$H_{\theta }$, and $H_{\varphi }$ from (19), (20), and (21), respectively.
Moreover, if we substitute $U\left( r\right) =R\left( r\right) /r$ in (19)
then $U\left( 0\right) =0=U\left( \infty \right) $. Yet, whilst $\Theta
\left( 0\right) $ and $\Theta \left( \pi \right) $ should be finite, $\Phi
\left( \varphi \right) $ should satisfy the single-valuedness condition $%
\Phi \left( \varphi \right) =\Phi \left( \varphi +2\pi \right) $. At this
point, we argue that the reality of the spectrum of Dirac eigenvalue
equation (6) is ensured not only by requiring $m,\Lambda ,\lambda \in 
%TCIMACRO{\U{211d} }%
%BeginExpansion
\mathbb{R}
%EndExpansion
$ but also by requiring $%
%TCIMACRO{\U{211d} }%
%BeginExpansion
\mathbb{R}
%EndExpansion
\ni \lambda +M^{2}=E^{2}>0$. With this understanding, we may now seek some $%
\mathcal{PT}$-symmetrization recipe (be it L\'{e}vai's [37] regular $%
\mathcal{PT}$-symmetrization or $\mathcal{P}_{\varphi }\mathcal{T}_{\mathcal{%
\varphi }}$-symmetrization of Mazharimousavi [36]) for each (at a time) of
the descendant Hamiltonians in (26).

\section{Consequences of complexified $\mathcal{P}_{\protect\varphi }%
\mathcal{T}_{\protect\varphi }$-symmetrized azimuthal Hamiltonians}

The eigenvalue equation in (21) with $a\in 
%TCIMACRO{\U{211d} }%
%BeginExpansion
\mathbb{R}
%EndExpansion
$ as a coupling parameter in a complexified-azimuthal effective interaction
of the form%
\begin{equation}
V_{eff}\left( \varphi \right) =-a^{2}e^{i\varphi },
\end{equation}%
would read$_{{}}$%
\begin{equation}
\left[ \frac{d^{2}}{d\varphi ^{2}}+a^{2}e^{i\varphi }+m^{2}\right] \Phi
\left( \varphi \right) =0.
\end{equation}%
Hence, a change of variable of the form $\ z=e^{i\varphi /2}$ would result in

\begin{equation}
z^{2}\frac{d^{2}\Phi \left( z\right) }{dz^{2}}+z\frac{d\Phi \left( z\right) 
}{dz}-\left( 4m^{2}+4a^{2}z^{2}\right) \Phi \left( z\right) =0.
\end{equation}%
Obviously, equation (29) is the modified Bessel equation with imaginary
argument and has two independent solutions. The linear combination of which
reads the general solution%
\begin{equation*}
\Phi \left( z\right) =C_{1}I_{2m}\left( 2az\right) +C_{2}K_{2m}\left(
2az\right) .
\end{equation*}%
Each of these independent solutions should identically satisfy the
single-valuedness condition $\Phi \left( \varphi \right) =\Phi \left(
\varphi +2\pi \right) $. One may, nevertheless, use the identities of [48]
and closely follow Mazharimousavi's treatment (namely, equations (17) - (28)
in [36]) and show that%
\begin{equation}
I_{2m}\left( 2ae^{i\left( \varphi +2\pi \right) /2}\right) =I_{2m}\left(
2ae^{i\varphi /2}\right) ,
\end{equation}%
\begin{equation}
K_{2m}\left( 2ae^{i\left( \varphi +2\pi \right) /2}\right) \neq K_{2m}\left(
2ae^{i\varphi /2}\right) .
\end{equation}%
Therefore, the regular solution collapses into%
\begin{equation}
\Phi \left( z\right) =C_{1}I_{2m}\left( 2az\right) \Longrightarrow \Phi
\left( \varphi \right) =C_{1}I_{2m}\left( 2ae^{i\varphi /2}\right) ,
\end{equation}

Under such settings, it is obvious that the Hamiltonian represented in (28)
reads%
\begin{equation}
H_{\varphi }=-\frac{d^{2}}{d\varphi ^{2}}-a^{2}e^{i\varphi /2},
\end{equation}%
and qualifies to be a $\mathcal{P}_{\varphi }\mathcal{T}_{\varphi }$%
-symmetric non-Hermitian Hamiltonian. That is, 
\begin{equation*}
\mathcal{P}_{\varphi }\mathcal{T}_{\varphi }H_{\varphi }\mathcal{P}_{\varphi
}\mathcal{T}_{\varphi }=H_{\varphi }.
\end{equation*}%
On the other hand, the eigenvalue equation (21) with $H_{\varphi }$ would
admit \ a regular azimuthal solution represented by the modified Bessel
function%
\begin{equation}
\Phi \left( \varphi \right) =C_{m,a}I_{2m}\left( 2ae^{i\varphi /2}\right) 
\text{ };\text{ }m=0,\pm 1,\pm 2,\cdots .
\end{equation}%
and satisfies the single-valuedness condition $\Phi \left( \varphi \right)
=\Phi \left( \varphi +2\pi \right) $ with $C_{m,a}$ as the normalization
constant to be found through the relation%
\begin{equation}
1=\left\langle \Phi _{m}\left( \varphi \right) /\mathcal{P}_{\varphi }%
\mathcal{T}_{\varphi }\Phi _{m}\left( \varphi \right) \right\rangle
=\left\vert C_{m,a}\right\vert ^{2}\int_{0}^{2\pi }\left\vert I_{2m}\left(
2ae^{i\varphi /2}\right) \right\vert ^{2}d\varphi .
\end{equation}%
Hereby, we have used the fact that our $\Phi _{m}\left( \varphi \right) $ in
(34) is $\mathcal{P}_{\varphi }\mathcal{T}_{\varphi }$-symmetric satisfying $%
\mathcal{P}_{\varphi }\mathcal{T}_{\varphi }\Phi _{m}\left( \varphi \right)
=\Phi _{m}\left( \varphi \right) $.

This would, in effect, suggest that since $%
%TCIMACRO{\U{211d} }%
%BeginExpansion
\mathbb{R}
%EndExpansion
\ni m=0,\pm 1,\pm 2,\cdots $ and $H_{\theta }$ of (20) is therefore
Hermitian, then $H_{\theta }$ of (20) admits real eigenvalues represented by 
$\Lambda \in 
%TCIMACRO{\U{211d} }%
%BeginExpansion
\mathbb{R}
%EndExpansion
$. Some illustrative consequences (with $H_{\theta }$ of (20) kept
Hermitian) are in order.

\subsection{Consequences of $V\left( \protect\theta \right) =0$ in (23)}

Should $V\left( \theta \right) =0$, one may clearly observe that equation
(20) is the very well known associated Legendre equation in which $\Lambda
=\ell \left( \ell +1\right) $, where $\ell $ is the angular momentum quantum
number, and $\Theta \left( \theta \right) =P_{\ell }^{m}(\cos \theta )$ are
the associated Legendre functions. Hence, following the regular textbook
procedure one may, in a straightforward manner, show that $\ell \geqslant
\left\vert m\right\vert $ (i.e., $m=0,\pm 1,\pm 2,\cdots ,\pm \ell $, is the
regular magnetic quantum number).

Consequently, as long as the Hermitian radial equation (19) is solvable
(could be exactly-, quasi-exactly-, conditionally-exactly-solvable, etc) for
the radial interaction $V_{eff}\left( r\right) $, the spectrum remains
invariant and real. However, the global wavefunction%
\begin{align}
\chi _{1}\left( r,\theta ,\varphi \right) & =\psi _{Sch}\left( r,\theta
,\varphi \right)  \notag \\
& =\sqrt{\frac{\left( 2l+1\right) \left( l-\left\vert m\right\vert \right) !%
}{2\left( l+\left\vert m\right\vert \right) !}}C_{ma}R_{n_{r}l}\left(
r\right) P_{l}^{m}\left( \cos \theta \right) I_{2m}\left( 2ae^{i\varphi
/2}\right) ,
\end{align}%
(with $n_{r}=0,1,2,\cdots $ as the radial quantum number) would indulge some
new probabilistic interpretations. This is due to the replacement of the
regular spherical harmonics $Y_{\ell m}\left( \theta ,\varphi \right) $ part
(for the radially symmetric 3D-Hamitonians) by the new $\mathcal{P}_{\varphi
}\mathcal{T}_{\varphi }$-symmetric part $P_{\ell }^{m}(\cos \theta )\Phi
_{m}\left( \varphi \right) $ (defined above for our $\mathcal{P}_{\varphi }%
\mathcal{T}_{\varphi }$-symmetric non-Hermitian Hamiltonian model).

To see the effect of such a $\mathcal{P}_{\varphi }\mathcal{T}_{\varphi }$%
-symmetrization on the probability density, we consider a radial Coulombic
effective interaction $V_{eff}\left( r\right) =-1/r$ accompanied by an
azimuthal effective interaction $V_{eff}\left( \varphi \right)
=-a^{2}e^{i\varphi }$(an illustrative example of fundamental nature). In
figures 1 and 2 we plot the corresponding probability densities at different
values of the coupling parameter $a$ for the principle quantum numbers $n=1$
and $n=2$ for $\ell =0=m$. It is clearly observed that whilst the
probability density for small $a$ imitates the Hermitian $\varphi $%
-independent probability density trends, it shifts and intensifies about $%
\left\vert \varphi \right\vert =0$ as $a$ increases (indicating that the
corresponding state is more localized, therefore). In this case, of course,
the rotational symmetry of a purely "just-radially-symmetric" Coulombic
interaction breaks down as a result of $V_{eff}\left( \varphi \right) .$

\subsection{Consequences of $V\left( \protect\theta \right) =1/2$ or $%
V\left( \protect\theta \right) =1/\left( 2\cos ^{2}\protect\theta \right) $
in (23)}

Taking $V\left( \theta \right) =1/2$ in (23) would imply%
\begin{equation}
\left[ \frac{1}{\sin \theta }\frac{d}{d\theta }\left( \sin \theta \frac{d}{%
d\theta }\right) -\frac{\widetilde{m}^{2}}{\sin ^{2}\theta }+\Lambda \right]
\Theta \left( \theta \right) =0,
\end{equation}%
where $\widetilde{m}=\sqrt{E+M+m^{2}}$ for Dirac and $\widetilde{m}=\sqrt{%
1/2+m^{2}}$ for Schr\"{o}dinger settings. Similar equation was reported by
Dutra and Hott [47]. The regular solution of which can (taking $\alpha
=\beta =0$ and $\gamma =1$ in equation (12) of ref. [47] to match with our
settings) very well be copied and pasted to read (for Dirac equation)%
\begin{equation}
\Theta \left( \theta \right) =y^{\rho }\left( 1-y\right) ^{\upsilon
}{}_{2}F_{1}\left( -k,b;d;y\right) \text{ ; \ }y=\cos ^{2}\left( \theta
/2\right) ,
\end{equation}%
where%
\begin{equation}
\upsilon =\rho =\frac{1}{2}\sqrt{m^{2}+E+M}\text{ };\text{ }b=k+4\upsilon +1;%
\text{ }d=1+2\upsilon ,
\end{equation}%
$k=0,1,2,\cdots $ is a "new" quantum number, and%
\begin{equation}
\Lambda =\frac{1}{4}\left( b+k\right) ^{2}-\frac{1}{4}=\frac{1}{4}\left[
\left( b+k+1\right) \left( b+k-1\right) \right] .
\end{equation}

On the other hand, $V\left( \theta \right) =1/\left( 2\cos ^{2}\theta
\right) $ would (taking $\alpha =\beta =0$ and $\gamma =1$ in equation (13)
of ref.[12) for Dirac equation) result in 
\begin{equation}
\rho =\frac{1}{4}+\frac{1}{4}\sqrt{1+4\left( E+M\right) };\text{ }\upsilon =%
\frac{1}{2}\sqrt{m^{2}+E+M}\text{\ }
\end{equation}%
\begin{equation}
b=k+2\left( \rho +\upsilon \right) +\frac{1}{2};\text{ }d=2\rho +\frac{1}{2}
\end{equation}%
and%
\begin{equation}
\Lambda =\left( b+k\right) ^{2}-\frac{1}{4}=\left[ \left( b+k+\frac{1}{2}%
\right) \left( b+k-\frac{1}{2}\right) \right]
\end{equation}%
For Schr\"{o}dinger case, nevertheless, one may just replace the term $%
\left( E+M\right) $ by $\left( 1/2\right) $ in the above expressions and get
the corresponding eigenvalue results. Then the general solution for both
cases would read%
\begin{align}
\chi _{1}\left( r,\theta ,\varphi \right) & =\psi _{Sch}\left( r,\theta
,\varphi \right)  \notag \\
& =N_{n_{r},k,m}R_{n_{r},k}\left( r\right) y^{\rho }\left( 1-y\right)
^{\upsilon }{}_{2}F_{1}\left( -k,b;d;y\right) I_{2m}\left( 2ae^{i\varphi
/2}\right) ,
\end{align}%
where $N_{n_{r},k,m}$ is the normalization constant that can be obtained in
a straightforward textbook procedure. Hereby, we witness that the general
solution (44) exhibits the change not only in the azimuthal part but also in
the angular $\theta $-part.

\section{Preservation of the magnetic quantum number $m$ and isospectrality}

To keep the magnetic quantum number as is (i.e., $m=0,\pm 1,\pm 2,\cdots $),
one may consider the azimuthal part of the general solution to be of the form%
\begin{equation}
\Phi _{m}\left( \varphi \right) =e^{im\varphi }F\left( \varphi \right) ,
\end{equation}%
where $F\left( \varphi \right) $ satisfies the single-valuedness condition $%
F\left( \varphi \right) =F\left( \varphi +2\pi \right) $.

Under such setting, the corresponding eigenvalue equation in (21) (with
primes denoting derivatives with respect to $\varphi $) reads%
\begin{equation}
F^{\prime \prime }\left( \varphi \right) +2imF^{\prime }\left( \varphi
\right) -V_{eff}\left( \varphi \right) F\left( \varphi \right) =0.
\end{equation}%
In this case $F\left( \varphi \right) $ would serve as a generating function
for the sought after azimuthal potential $V_{eff}\left( \varphi \right) $
and shapes the form of the azimuthal solution $\Phi _{m}\left( \varphi
\right) $. As an illustrative example, a generating function $F\left(
\varphi \right) =\cos \varphi $ would imply%
\begin{equation}
V_{eff}\left( \varphi \right) =-\left[ 1+2im\tan \varphi \right]
\end{equation}%
which is indeed a non-Hermitian and $\mathcal{P}_{\varphi }\mathcal{T}%
_{\varphi }$-symmetric, $\mathcal{P}_{\varphi }\mathcal{T}_{\varphi
}V_{eff}\left( \varphi \right) =V_{eff}\left( \varphi \right) $.

However, one may wish to follow the other way around and consider a $%
\mathcal{P}_{\varphi }\mathcal{T}_{\varphi }$-symmetric $V_{eff}\left(
\varphi \right) $ and solve (46) for $F\left( \varphi \right) $. In this
manner, $V_{eff}\left( \varphi \right) $ would now serve as a generating
function for $F\left( \varphi \right) $ and consequently a generating
function for $\Phi _{m}\left( \varphi \right) $. An immediate example is in
order. Consider%
\begin{equation}
V_{eff}\left( \varphi \right) =-\frac{\omega ^{2}}{4}e^{i\varphi }
\end{equation}%
and solve (44) for a regular $F\left( \varphi \right) $ to obtain%
\begin{equation}
F\left( \varphi \right) =C_{\circ }e^{-im\varphi }I_{2m}\left( \omega
e^{i\varphi /2}\right) ;\text{ }%
%TCIMACRO{\U{211d} }%
%BeginExpansion
\mathbb{R}
%EndExpansion
\ni m=0,\pm 1,\pm 2,\cdots \text{.}
\end{equation}%
Then, (45) would read%
\begin{equation}
\Phi _{m}\left( \varphi \right) =C_{m,\omega }I_{2m}\left( \omega
e^{i\varphi /2}\right)
\end{equation}%
It is, therefore, obvious that all effective potentials $V_{eff}\left(
\varphi \right) $ satisfying (48) would essentially change the azimuthal
part of the general solution.

Moreover, in the process of preserving the magnetic quantum number $m$ as
defined in the regular Hermitian settings, a set of isospectral $\varphi $%
-dependent potentials, $V_{eff}\left( \varphi \right) $, is obtained. That
is, for each set of $V_{eff}\left( r\right) ,V_{eff}\left( \theta \right)
\in 
%TCIMACRO{\U{211d} }%
%BeginExpansion
\mathbb{R}
%EndExpansion
$, all $\varphi $-dependent potentials, $V_{eff}\left( \varphi \right) $,
satisfying (46) are isospectral.

\section{Concluding remarks}

In the build up of a generalized quantum recipe (Bender's and Boettcher's
PTQM in this case), a question of delicate nature arises in the process as
to "would PTQM be able to recover some results (if not all, to be classified
as a promising theory) of the conventional Hermitian quantum mechanics?". To
the best of our knowledge, only rarely and mainly within regular Hermitian
(but $\mathcal{PT}$-symmetric) settings examples were provided such as the
one by Bender, Brody and Jones [35] mentioned in our introduction section
above (i.e., $H=p^{2}+x^{2}+2x$). The reality of the energy eigenvalues and
other quantum mechanical properties (rather than the "recoverability of
Hermitian quantum mechanical" results) were the main constituents and focal
points in the studies of the non-Hermitian $\mathcal{PT}$-symmetric
Hamiltonians. In our current proposal, with a new class of non-Hermitian $%
\mathcal{P}_{\varphi }\mathcal{T}_{\varphi }$-symmetric Hamiltonians (having
real spectra identical to their Hermitian partner Hamiltonians), we tried to
fill this gap, at least partially.

Through our over simplified non-Hermitian $\mathcal{P}_{\varphi }\mathcal{T}%
_{\varphi }$-symmetrized Hamiltonian (33), we have shown that some
conventional relativistic and non-relativistic quantum mechanical results
are indeed recoverable (the energy eigenvalues here). We have witnessed,
however, an unavoidable change in the azimuthal part of the general
wavefunction. Such a change would introduce some new probabilistic
interpretations. With $V\left( \theta \right) =0$ and $V_{eff}\left( \varphi
\right) =-a^{2}e^{i\varphi }$, for example, we have observed that a quantum
state becomes more localized as the probability density intensifies at a
specific point $\left\vert \varphi \right\vert =0$ (documented in figures 1
and 2 of section 3.1). Moreover, setting $V\left( \theta \right) \neq 0$
(again with $V_{eff}\left( \varphi \right) =-a^{2}e^{i\varphi }$ in $%
H_{\varphi }$) in the descendant Hamiltonian $H_{\theta }$ has indeed
manifested a change in the angular $\theta $-dependent part of the general
solution too (documented in section 3.2). This would, of course, has some
"new" effects on the probabilistic interpretations in turn. Yet, a recipe to
keep the magnetic quantum number $m$ as defined in the regular Hermitian
quantum mechanical settings is suggested.

In connection with the current proposal's spherical-separability and
non-Hermiticity, it is obvious that the descendant Hamiltonians $H_{r}$, $%
H_{\theta },$ and $H_{\varphi }$ play essential roles and offer some
"user-friendly", say, options as to which one (or ones) of them is (or are)
non-Hermitian. Be it $\mathcal{P}_{\varphi }\mathcal{T}_{\varphi }$%
-symmetric, $\mathcal{PT}$-symmetric, pseudo-Hermitian or $\eta $%
-pseudo-Hermitian, they very well fit into Bender's and Boettcher's PTQM
(irrespective with their nicknames and with the understanding that $\mathcal{%
P}$ and $\mathcal{T}$ need not necessarily identify just parity and time
reversal, respectively). Yet, a complexification of $0\neq V\left( \theta
\right) \in 
%TCIMACRO{\U{2102} }%
%BeginExpansion
\mathbb{C}
%EndExpansion
$ in $H_{\theta }$ with the understanding that a parity-like $\mathcal{P}%
_{\theta }$ and a time reversal-like $\mathcal{T}_{\theta }$ operators may
very well suggest a similar $\mathcal{P}_{\theta }\mathcal{T}_{\theta }$%
-symmetric $H_{\theta }$ Hamiltonian. Such non-Hermitian $\mathcal{P}%
_{\varphi }\mathcal{T}_{\varphi }$-symmetrized and/or $\mathcal{P}_{\theta }%
\mathcal{T}_{\theta }$-symmetrized (anticipated to be feasible but yet to be
identified) Hamiltonians better be nicknamed as \emph{pseudo-}$\mathcal{PT}$%
\emph{-symmetric Hamiltonians.\newpage }

\newpage

\textbf{Figures' Captions:}

\textbf{Figure 1:} Shows the effect of $V_{eff}\left( \varphi \right)
=-a^{2}e^{i\varphi }$ on the probability density as the coupling parameter $%
a $ increases for $n=1,\ell =0=m$.

\textbf{Figure 2:} Shows the effect of $V_{eff}\left( \varphi \right)
=-a^{2}e^{i\varphi }$ on the probability density as the coupling parameter $%
a $ increases for $n=2,\ell =0=m$.


\begin{thebibliography}{99}
\bibitem{} C M Bender and S Boettcher, Phys. Rev. Lett. {\ }\textbf{80}
(1998) 5243

\bibitem{} C M Bender, S Boettcher and P N Meisinger, J. Math. Phys. \textbf{%
40} (1999) 2201

\bibitem{} B Bagchi, F Cannata and C Quesne, Phys. Lett. \textbf{A 269}
(2000) 79

\bibitem{} Z Ahmed, Phys. Lett. \textbf{A 282 }(2001)\textbf{\ }343

\bibitem{} Z Ahmed, Phys. Lett. \textbf{A 287 }(2001)\textbf{\ }295

\bibitem{} Z Ahmed, Phys. Lett. \textbf{A 290 }(2001)19

\bibitem{} A Khare and B P Mandal, Phys. Lett. \textbf{A 272 }(2000)\textbf{%
\ }53

\bibitem{} V Buslaev and V Grecchi, J. Phys.{\ \textbf{A}: Math. Gen. }%
\textbf{26} (1993) 5541

\bibitem{} M Znojil and G L\'{e}vai, Phys. Lett. A \textbf{271} (2000) 327

\bibitem{} B Bagchi, S Mallik, C Quesne and R Roychoudhury, Phys. Lett. 
\textbf{A 289} (2001) 34

\bibitem{} P Dorey, C Dunning and R Tateo, J. Phys. \textbf{A}: Math. Gen. 
\textbf{34} (2001) 5679

\bibitem{} R Kretschmer and L Szymanowski, Czech. J.Phys \textbf{54} (2004)
71

\bibitem{} M Znojil, F Gemperle and O Mustafa, J. Phys. \textbf{A}: Math.
Gen. \textbf{35} (2002) 5781

\bibitem{} O Mustafa and M Znojil, J. Phys. \textbf{A}: Math. Gen. \textbf{35%
} (2002) 8929

\bibitem{} B F Samsonov and P Roy, J. Phys. \textbf{A}: Math. Gen. \textbf{38%
} (2005) L249.

\bibitem{} A Mostafazadeh, J. Math. Phys. \textbf{43 }(2002) 2814

\bibitem{} A Mostafazadeh, Nucl.Phys. \textbf{B 640} (2002) 419

\bibitem{} A Mostafazadeh, J. Math. Phys. \textbf{43 }(2002) 205

\bibitem{} A Mostafazadeh, J. Math. Phys. \textbf{43} (2002) 3944

\bibitem{} A Mostafazadeh, J. Math. Phys. \textbf{44 }(2003) 974

\bibitem{} A Mostafazadeh, J. Phys. \textbf{A}: Math. Gen. \textbf{38}
(2005) 3213

\bibitem{} A Sinha and P Roy, Czech. J. Phys. \textbf{54} (2004) 129

\bibitem{} L Jiang, L Z Yi and C S Jia, Phys Lett \textbf{A 345} (2005) 279

\bibitem{} B P Mandal, Mod. Phys. Lett. \textbf{A 20 }(2005) 655

\bibitem{} B Znojil, H B\'{\i}la and V Jakubsky, Czech. J. Phys. \textbf{54}
(2004) 1143

\bibitem{} A Mostafazadeh and A Batal, J. Phys.\textbf{A}: Math. Gen. 
\textbf{37 }(2004) 11645

\bibitem{} B Bagchi and C Quesne, Phys. Lett. \textbf{A 301} (2002) 173

\bibitem{} L Solombrino, J. Math. Phys. \textbf{43} (2002) 5439

\bibitem{} T V Fityo, J. Phys. \textbf{A}: Math. Gen. \textbf{35} (2002) 5893

\bibitem{} O Mustafa and S H Mazharimousavi, Czech. J. Phys. \textbf{56 }%
(2006) 967 (arXiv: quant-ph/0603237)

\bibitem{} O Mustafa and S H Mazharimousavi, Phys. Lett. \textbf{A 357}
(2006) 295 (arXiv: quant-ph/0604106)

\bibitem{} O Mustafa and S H Mazharimousavi, Int. J. Theor. Phys., (2008) in
press (arXiv: hep-th/0601017)

\bibitem{} B Bagchi and C Quesne, Phys. Lett. \textbf{A 301} (2002) 173

\bibitem{} M Znojil, "$\mathcal{PT}$-symmetry, ghosts, supersymmetry and
Klien-Gordon equation", (2004) arXiv: quant-ph/0408081

\bibitem{} C M Bender, D C Brody and H F Jones, Am. J. Phys. \textbf{71}
(2003) 1095

\bibitem{} S H Mazharimousavi, J. Phys. \textbf{A}; Math. Theor., (2008) in
press (arXiv: quant-ph/0801.1549)

\bibitem{} G L\'{e}vai, J. Phys. \textbf{A}: Math. Theor. \textbf{40} (2007)
F273

\bibitem{} H Hartmann, Theor. Chim. Acta \textbf{24} (1972) 201.

\bibitem{} H Hartmann, R Schuck, and J Radtke, Theor. Chim. Acta \textbf{46}
(1976) 1.

\bibitem{} H Hartmann and D Schuck, Int. J. Quantum Chem. \textbf{18} (1980)
125.

\bibitem{} C\ C Gerry, Phys. Lett. \textbf{A 118} (1986) 445.

\bibitem{} C Y Chen and S H Dong, Phys. Lett. \textbf{A 335} (2005) 374.

\bibitem{} C Y Chen, F L Lu, and D S Sun, Phys. Lett. \textbf{A 329} (2004)
420.

\bibitem{} C Y Chen, D S Sun and C L Liu, Phys. Lett. \textbf{A 317} (2003)
80.

\bibitem{} C Y Chen, C L Liu and D S Sun, Phys. Lett. \textbf{A 305} (2002)
341.

\bibitem{} C Y Chen, Phys. Lett. \textbf{A 339} (2005) 283.

\bibitem{} A de Souza Dutra and M Hott, Phys. Lett. \textbf{A 356} (2006)
215.

\bibitem{} Z X Wang and D R Guo, "Special Functions" (Singapore: World
Scientific), (1989) page 428.
\end{thebibliography}
\end{document}